\documentclass[10pt,conference,letterpaper]{IEEEtran}
\usepackage{cite}
\usepackage{graphicx}
\usepackage[percent]{overpic}
\usepackage{amsmath}
\usepackage{amssymb}
\usepackage{textcomp}
\usepackage{algorithm}
\usepackage{algpseudocode}
\usepackage{booktabs}
\usepackage{makecell}
\usepackage{url}

\title{Parallelizing Large-Scale Tensor Network Contraction on Multiple GPUs}

\begin{document}

\IEEEoverridecommandlockouts

\author{\IEEEauthorblockN{Feng Pan$^{\ast}$\thanks{$^{\ast}$Both authors contributed equally to this research. Corresponding authors: F. Pan (feng\_pan@sutd.edu.sg) and H. Gu (henrygu@nvidia.com).}}
\IEEEauthorblockA{Singapore University \\
of Technology and Design}
\and
\IEEEauthorblockN{Hanfeng Gu$^{\ast}$}
\IEEEauthorblockA{NVIDIA Corporation}
\and
\IEEEauthorblockN{Paul Springer}
\IEEEauthorblockA{NVIDIA Corporation}
\and
\IEEEauthorblockN{Xipeng Li}
\IEEEauthorblockA{NVIDIA Corporation}}

\maketitle

\begin{abstract}
  Exact tensor network contraction underpins quantum circuit simulation, quantum error correction, combinatorial optimization, and many-body dynamics. The dominant parallelization strategy, slicing, scales exponentially and incurs redundant computation. We present a multi-GPU framework that instead distributes intermediate tensors across devices with explicit communication, converting a fixed contraction path into a communication-efficient schedule via GEMM-oriented mode reordering and communication-aware mode distribution planning. Within a single DGX H100 node (8 GPUs, NVLink), distribution delivers $7$--$173\times$ extra speedup beyond embarrassingly parallel slicing, capturing nearly all of the available compute reduction (87--101\%) because NVLink's high bandwidth keeps communication small relative to compute. Scaling the same four workloads to 1024 H100 GPUs over InfiniBand, the extra speedup beyond slicing ranges from $42\times$ to $67{,}869\times$, demonstrating that communication-aware distributed contraction far surpasses slicing-based scaling limits for frontier tensor networks.
\end{abstract}


\section{Introduction}
Classical simulation of high-dimensional systems remains a central challenge in computational science.
Tensor networks (TNs) address this challenge by factorizing a high-order tensor into a structured network of smaller tensors that can be stored and manipulated much more efficiently~\cite{schollwock2011dmrg, orus2014practical}.
Their importance is no longer limited to condensed-matter physics.
They now form a broadly useful computational language for exact and approximate algorithms in quantum circuit simulation, quantum error correction, combinatorial optimization, quantum many-body dynamics, and tensorized machine learning~\cite{gray2021hyper, stoudenmire2016supervised, novikov2015tensorizing, ferris2014tensor, liu2021tropical, fauseweh2024manybody}.
Across these domains, the same core operation appears repeatedly: TN contraction, i.e., summing over shared indices to produce amplitudes, coset probabilities, counting results, partition functions, or other observables of interest.

The need for scalable TN contraction is increasingly urgent.
Random circuit sampling continues to test the classical boundary of quantum advantage, logical-qubit experiments demand stronger exact decoding baselines, exact counting problems on irregular graphs remain computationally difficult, and quantum hardware is beginning to probe many-body dynamical regimes that are hard to verify classically~\cite{arute2019quantum, google2023suppressing, liang2025kings, king2025beyondclassical}.
For large instances, however, contraction rapidly becomes the dominant bottleneck.
Its cost depends acutely on the contraction path: different orders of pairwise contractions can lead to orders-of-magnitude differences in floating-point work, memory traffic, and peak intermediate size.
Finding a globally optimal path is NP-hard in general~\cite{markov2008simulating}, so practical software relies on high-quality heuristics together with memory-management techniques such as slicing and path reconfiguration~\cite{gray2021hyper}.
Even then, frontier-scale problems often remain limited not by arithmetic throughput but by memory capacity.
Once an intermediate tensor no longer fits in the high-bandwidth memory of a single GPU, the contraction simply becomes infeasible.

This paper focuses on that systems-level barrier.
Our goal is not only to accelerate tensor contractions on multiple GPUs, but to make previously infeasible contractions executable by combining communication-aware distribution with tensor-aware data layout.
The central idea is that multi-GPU execution should be decided using the structure of the contraction path itself: which modes are kept, which are reduced, where large intermediate tensors appear, and where redistribution can avoid expensive communication later in the path.

This paper makes three main contributions.
First, we present a systematic framework for converting a fixed TN contraction path into a multi-GPU execution schedule, explicitly addressing data partitioning, communication, and execution ordering.
Second, we introduce an offline scheduling strategy that combines GEMM-oriented mode reordering with communication-aware mode distribution planning; a dynamic-programming search guided by a hardware-aware cost model selects partitioned modes and redistribution points for a target GPU count.
Third, we evaluate the framework experimentally on quantum circuit simulation and three many-body dynamics benchmarks, demonstrating generality across application families by scaling all four workloads from a single NVLink node to 1024 GPUs over InfiniBand.

\section{Background}
\subsection{Tensor Networks}
TNs represent a large tensor as a collection of lower-order tensors connected by shared indices~\cite{schollwock2011dmrg, orus2014practical}.
This factorized view is especially powerful when the target problem has locality, limited entanglement, or other structure that allows low-rank compression.
In practice, the same formalism supports both approximate methods, such as matrix product state time evolution, and exact contraction algorithms used in quantum circuit simulation and counting problems.

A TN can be viewed as a graph whose nodes are tensors (multidimensional arrays) and whose edges denote shared indices.
\emph{Closed} indices connect two tensors and are summed over during contraction; \emph{open} indices survive in the final output.
We focus on exact contraction: summing over all closed indices while preserving the open ones.
For two tensors, $C_{ij} = \sum_k A_{ik} B_{kj}$; larger networks apply the same rule repeatedly.

\subsection{Contraction Path and Complexity}

In most implementations, a TN is evaluated as a sequence of pairwise contractions.
Although the final result is independent of the contraction order, the computational cost is not.
This ordered sequence is called the \emph{contraction path} and can be represented as a binary contraction tree whose leaves are the original tensors and whose internal nodes are intermediate tensors.

The quality of a contraction path is usually measured by three related metrics:
\begin{enumerate}
    \item \textbf{Time complexity} ($C_t$): the total number of floating-point operations (FLOPs) required by the full contraction.
    \item \textbf{Space complexity} ($C_s$): the size of the largest intermediate tensor that appears along the path.
    \item \textbf{Memory complexity} ($C_m$): a proxy for total data movement, obtained by summing the sizes of the tensors read and written at each contraction step.
\end{enumerate}

Optimizing a contraction path for general TNs is NP-hard~\cite{markov2008simulating}.
Exhaustive search is therefore limited to very small instances, and practical software relies on heuristics, hypergraph partitioning, local tree rewrites, or stochastic search to obtain high-quality paths~\cite{gray2021hyper}.

To define these costs, consider one pairwise contraction step $a$ written as
\[
    A_{MK} \times B_{KN} \rightarrow C_{MN},
\]
where $M$ and $N$ denote the retained modes and $K$ denotes the reduced modes.
Let
\[
    m = \prod_{x \in M} d_x,\qquad
    n = \prod_{x \in N} d_x,\qquad
    k = \prod_{x \in K} d_x.
\]
The contraction-operation count of this step is then
\[
    C_t^{(a)} = mnk,
\]
the corresponding peak space requirement is
\[
    C_s^{(a)} = \max(mk,\; kn,\; mn),
\]
and the memory complexity is
\[
    C_m^{(a)} = mk + kn + mn.
\]

For a full contraction path, the total time and memory complexities are additive over steps,
\begin{equation}
    C_t = \sum_a C_t^{(a)}
\end{equation}
\begin{equation}
    C_m = \sum_a C_m^{(a)}
\end{equation}
whereas the space complexity is determined by the largest intermediate tensor encountered anywhere along the path:
\begin{equation}
    C_s = \max_a C_s^{(a)}.
\end{equation}
On modern accelerators, $C_s$ often determines feasibility, while $C_t$ and $C_m$ determine performance once the contraction fits in memory.
For clarity, the formulas above omit the constant factor associated with complex arithmetic.
In the experimental FLOP and throughput numbers, tensors use complex64 arithmetic and one complex multiply-add is counted as eight real FP32 FLOPs (four multiplies and four additions), matching the operation counter used by the path logs.

\subsection{Contraction Techniques}

Even a good contraction path may still require an intermediate tensor larger than device memory.
In that case, a standard remedy is \emph{slicing}: one or more indices are fixed to specific values, the resulting smaller contractions are executed independently, and the partial results are summed at the end.
Slicing reduces peak memory at the cost of repeating work across many subproblems.

State-of-the-art optimizers such as \texttt{cotengra}~\cite{gray2021hyper} interleave slicing with path refinement (dynamic slicing), substantially reducing overhead.
However, slicing does not remove the underlying source of hardness: contraction complexity grows exponentially with the treewidth of the network graph~\cite{markov2008simulating}, and the largest intermediate tensor is also exponential in treewidth.

For such large-treewidth problems, multi-GPU execution becomes more than a speed optimization.
It is often the only way to overcome the single-device memory wall while preserving exactness.
The challenge then shifts from path optimization alone to the coordinated design of data layout, communication, and redistribution across devices, which is the focus of this paper.

\subsection{Multi-GPU Hardware Platform}
\label{sec:hardware}

We evaluate on the NVIDIA DGX H100~\cite{nvidia_h100} (Table~\ref{tab:hw}): eight H100 GPUs per node, each with 80\,GB HBM3, interconnected by NVLink (900\,GB/s per GPU) within a node and InfiniBand (400\,Gb/s per GPU) across nodes.
This two-tier hierarchy---fast intra-node NVLink vs.\ slower inter-node InfiniBand---motivates our emphasis on limiting communication volume and avoiding fine-grained transfers.

\begin{table}[htb]
  \centering
  \caption{Key hardware parameters of the DGX H100 platform.
  All bandwidth figures are bidirectional per GPU.}
  \label{tab:hw}
  \small
  \begin{tabular}{lc}
    \toprule
    & DGX H100 \\
    \midrule
    GPUs per NVLink domain         & 8 \\
    GPU memory (HBM)               & 80\,GB \\
    HBM bandwidth                  & 3.35\,TB/s \\
    Peak FP32 FLOP/s per GPU       & 67\,TFLOP/s \\
    Aggregate domain memory        & 640\,GB \\
    NVLink generation              & 4th \\
    NVLink bandwidth / GPU         & 900\,GB/s \\
    Inter-node interconnect        & 400\,Gb/s IB \\
    \bottomrule
  \end{tabular}
\end{table}

\subsection{cuTENSORMp: Distributed Tensor Contraction Engine}
\label{sec:cutensormp}

Translating a multi-GPU distribution plan into efficient device-level execution requires coordinating local tensor arithmetic with inter-device data movement.
cuTENSORMp~\cite{cutensormp} is a multi-process extension of the cuTENSOR library that provides exactly this capability: given a distributed tensor contraction specified in Einstein summation notation, it automatically orchestrates local computation and inter-process communication across an arbitrary number of GPUs.

\subsubsection{Programming Model}
cuTENSORMp exposes a high-level interface in which the caller specifies:
\begin{enumerate}
  \item An einsum equation string defining the contraction, e.g., $\texttt{abcde,bdfg}\rightarrow\texttt{acefg}$.
  \item The global extents of each tensor mode.
  \item A per-mode \texttt{ranksPerMode} array for each operand and the output, indicating how many ranks share each mode.
  A mode with $\texttt{ranksPerMode}=1$ is replicated; a mode with $\texttt{ranksPerMode}=d_x$ is evenly partitioned across $d_x$ ranks along that dimension.
\end{enumerate}
From these descriptors the library constructs distributed tensor layouts, determines which data each rank owns, and builds an execution plan---all without requiring the caller to manage point-to-point transfers or collective schedules explicitly.

\subsubsection{Pipelined Computation--Communication Overlap}
For our purposes, the central systems feature of cuTENSORMp is its ability to \emph{overlap local computation with inter-device communication}.
Internally, the library decomposes a distributed contraction into a pipeline of three concurrent activities:
\begin{itemize}
  \item \textbf{Local compute.} Each GPU executes cuTENSOR tensor contraction kernels on its local shard of the operands.
  \item \textbf{Communication.} NCCL~\cite{nccl} collectives and point-to-point transfers redistribute, broadcast, or reduce tensor shards as dictated by the distribution layout.
  \item \textbf{Tiling and double-buffering.} Large contractions are decomposed into tiles; while one tile is being computed on the Streaming Multiprocessors, the results of the previous tile are simultaneously being transferred over NVLink or InfiniBand, and the operands for the next tile are being prefetched.
\end{itemize}
This pipelining is important for the workloads in this paper, where a single distributed contraction step may move tens of gigabytes of intermediate data.

\subsubsection{NCCL Transport and Network Awareness}
cuTENSORMp delegates all inter-process communication to NCCL, which automatically selects the best available transport: NVLink and NVSwitch for intra-node transfers and InfiniBand RDMA for cross-node links.

\subsubsection{Role in the Framework}
The availability of cuTENSORMp creates a clean separation of concerns in our system.
The offline planner (Section~\ref{sec:system}) decides \emph{which} modes to distribute, \emph{when} to redistribute, and \emph{when} to gather---a combinatorial scheduling problem.
cuTENSORMp then determines \emph{how} each scheduled contraction is realized on the target platform, including tiling, workspace management, and communication orchestration.
This separation lets the planner reason at the level of communication volume and transfer granularity, while the library maps those choices onto the available NVLink, NVSwitch, and InfiniBand substrate.

\subsection{Prior Work on Large-Scale TN Simulation}
\label{sec:related}

The rapid growth of quantum processors has driven a parallel effort in large-scale classical simulation using TNs.
We organize the most relevant prior work into three themes: contraction path optimization and slicing, supercomputer-scale quantum circuit simulation, and multi-GPU simulation systems.

\subsubsection{Contraction Path Optimization and Slicing}
At the algorithmic level, the performance of a TN simulator is largely determined by the contraction path.
Markov and Shi~\cite{markov2008simulating} connected simulation complexity to treewidth, providing the graph-theoretic basis for path search and width reduction.
More recently, path finders combine tree search, local rewrites, and dynamic slicing to find contraction schemes that trade peak memory for additional work.
Huang et al.~\cite{huang2021parallelization} showed that slicing can also expose substantial coarse-grained parallelism by decomposing a contraction into many independent subproblems, enabling more than $10^5\times$ acceleration on representative quantum-circuit benchmarks.
However, these techniques optimize the contraction itself; they do not by themselves define how an individual contraction step should be partitioned and executed efficiently across many GPUs.

\subsubsection{Supercomputer-Scale Quantum Circuit Simulation}
Random circuit sampling (RCS) has been the highest-profile target for distributed TN simulation.
Liu et al.~\cite{liu2021closing} demonstrated real-time Sunway-based simulation of the Sycamore task using three-level parallelism across 42 million cores, reporting one million correlated samples in 304 seconds.
Pan and Zhang~\cite{pan2022bigbatch} introduced the big-batch TN method for computing exact amplitudes or probabilities of large sets of correlated bitstrings, while Pan et al.~\cite{pan2022sampling} used 512 GPUs to generate one million uncorrelated samples for Sycamore circuits.
Zhao et al.~\cite{zhao2025leapfrogging} later used 1432 NVIDIA A100 GPUs to generate three million uncorrelated samples with an XEB comparable to Sycamore in 86.4 seconds.
Their implementation maps large stem tensors across 8 intra-node GPUs along the leading three dimensions and assigns sliced subtasks independently to nodes.
Fu et al.~\cite{fu2024surpassing} extended this line of work to 2304 GPUs at SC24, achieving a time-to-solution of 14.22 seconds through a system-level three-tier parallelization scheme with hybrid communication and low-precision quantization.
Chen et al.~\cite{chen2025swtnc} demonstrated the SW-TNC framework on Sunway, using data-reuse and step-fusion strategies to accelerate Zuchongzhi-60-24 simulation by over $10\times$ on 1024 nodes.

Viewed together, these RCS simulators rely primarily on slicing to expose coarse-grained parallelism and typically pair it with bespoke mappings of large intermediates onto fixed device hierarchies.
Our work instead targets the execution layer: we automatically choose which modes to partition, when to redistribute, and when to gather across an \emph{arbitrary} number of GPUs, and then realize those decisions through cuTENSORMp mode-distributed contractions with pipelined computation--communication overlap.

\subsubsection{Multi-GPU Simulation Systems and Libraries}
On the systems side, Atlas~\cite{xu2024atlas} performs Schr\"{o}dinger-style (state-vector) quantum circuit simulation on multi-GPU clusters using hierarchical partitioning and ILP-based staging to reduce communication.
In contrast to TN simulators, state-vector methods still scale with the full $2^n$ state size, even when extended with host-memory or DRAM offload, whereas TN contraction can exploit circuit structure to reach larger qubit counts.
Brown et al.~\cite{brown2025multigpu} benchmarked multi-GPU quantum simulation across several GPU generations and showed that interconnect advances have contributed more than $16\times$ improvements in time-to-solution.
At the library level, cuTensorNet provides high-performance TN contraction, including distributed execution built mainly around slice-based parallelism~\cite{cicero2025simulation}, whereas our framework builds on the complementary cuTENSORMp library to support mode-distributed contractions with explicit computation--communication overlap.

\section{Motivating Applications}
The need for distributed exact TN contraction extends well beyond quantum circuit simulation.
Fig.~\ref{fig:complexity-reduction} shows the compute-only complexity reduction (the FLOP reduction, ignoring communication) from distributing intermediates across GPUs for six workloads; in all cases, contraction cost drops rapidly as more devices contribute memory.
Section~\ref{sec:experiments} provides end-to-end measured results confirming that these reductions translate into real wall-clock gains.

\subsection{Quantum Circuit Simulations}
Random circuit sampling (RCS) is one of the most demanding benchmarks for exact TN contraction because it intentionally generates highly entangled networks with large effective treewidth.
As circuit depth increases, the computation becomes dominated by a small number of very large intermediates, making memory capacity and communication at least as important as arithmetic throughput.
This benchmark has become a focal point in the race between quantum hardware and classical simulation algorithms~\cite{arute2019quantum, wu2021strong, zlokapa2023boundaries}.
On the classical side, TN methods have progressed from approximate contraction on arbitrary networks to big-batch amplitude evaluation, direct sampling, and large-scale GPU and supercomputer implementations~\cite{pan2020contracting, pan2022bigbatch, pan2022sampling, pan2024gpus, huang2021parallelization, liu2021closing, fu2024surpassing, zhao2025leapfrogging}.
This is the primary experimental focus of our paper, because Sycamore- and Zuchongzhi-style circuits sit squarely in the regime where a good path alone is insufficient and efficient multi-GPU execution becomes decisive.
Fig.~\ref{fig:complexity-reduction}(a) shows the complexity reduction for Zuchongzhi n60m24 as the number of GPUs grows from 1 to 1024.

\subsection{Quantum Error Correction}
Quantum Error Correction (QEC) is essential for realizing fault-tolerant quantum computing.
Exact maximum-likelihood decoding can be written as a TN contraction over all error configurations consistent with an observed syndrome~\cite{bravyi2014efficient, ferris2014tensor, fowler2012surface}.
For surface codes under circuit-level noise, the resulting networks become effectively three-dimensional and quickly exceed single-GPU memory as code distance grows.
This matters increasingly in practice because logical-qubit experiments and learned decoders both require strong exact baselines~\cite{piveteau2024beyond2d, google2023suppressing, bausch2024learning, google2025belowthreshold}.
Fig.~\ref{fig:complexity-reduction}(b) shows that exact distance-$7$ rotated-surface-code decoding benefits strongly from distributed TN contraction, with contraction cost dropping rapidly as more GPUs participate.

\subsection{Combinatorial Optimization Problems}
Many counting and optimization problems can be reformulated as exact TN contractions by encoding local constraints into small tensors.
These workloads are especially useful because they often generate irregular contraction trees and nonuniform communication patterns rather than the regular structures seen in lattice physics.
Tropical and semiring formulations, including recent work on independent-set enumeration on King's graphs, show that exact TN methods can push well beyond naive search~\cite{liu2021tropical, liang2025kings}.
Fig.~\ref{fig:complexity-reduction}(c) shows that this irregular workload also benefits substantially from distributed contraction, reinforcing that the need for multi-GPU execution is not limited to circuit-like networks.

\subsection{Simulation of Quantum Many-Body Dynamics}
\begin{figure*}[t]
    \centering
    \includegraphics[width=0.92\linewidth]{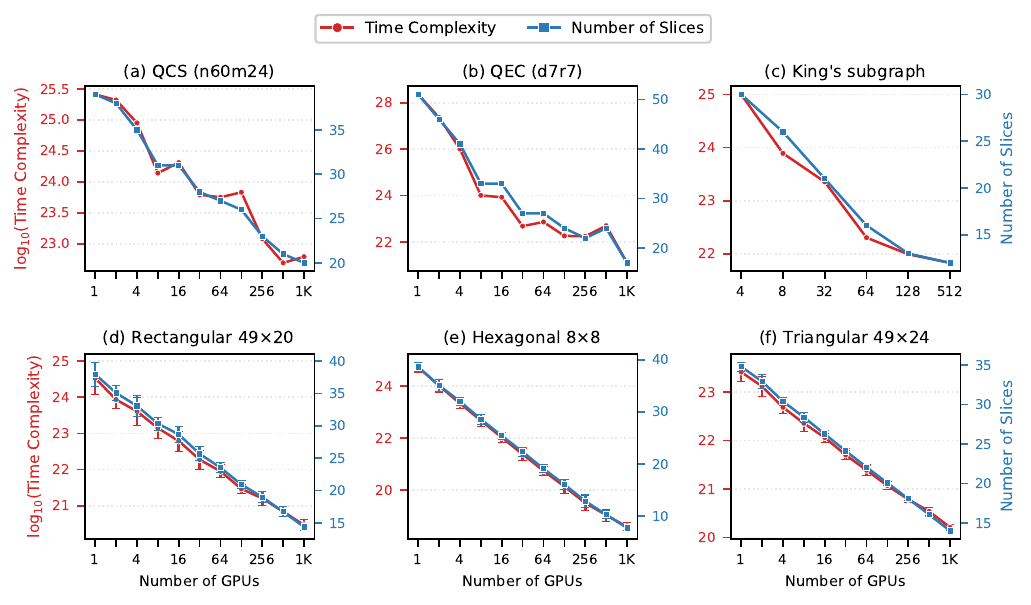}
    \caption{Theoretical complexity reduction from distributing intermediate tensors across GPUs for six workloads:
    (a)~quantum circuit simulation (Zuchongzhi n60m24),
    (b)~exact QEC decoding (distance-7 rotated surface code),
    (c)~independent-set enumeration on King's subgraphs,
    and many-body dynamics on (d)~rectangular, (e)~hexagonal, and (f)~triangular lattices.
    Red curves show log$_{10}$(total FLOPs); blue curves show the number of sliced indices.
    In all cases, contraction cost drops rapidly---often close to exponentially---as more GPUs contribute shared memory.
    Many-body panels show mean $\pm$ std over 15 independent path-finding runs.}
    \label{fig:complexity-reduction}
\end{figure*}

Real-time quantum dynamics can also be mapped to deep spacetime TNs through Suzuki-Trotter decomposition and related time-evolution methods~\cite{suzuki1976generalized, vidal2004efficient}.
In two-dimensional lattices or long-time simulations, bond dimensions and intermediate tensors grow rapidly, so access to distributed high-bandwidth memory becomes just as important as local compute throughput.
This is especially timely because IBM, D-Wave, and Google experiments are already probing many-body or gauge-theory regimes that are increasingly difficult to verify classically~\cite{fauseweh2024manybody, kim2023utility, shtanko2025integrability, king2023spinglass, king2025beyondclassical, mildenberger2025confinement}.
Fig.~\ref{fig:complexity-reduction}(d--f) show the same rapid decrease in contraction cost across rectangular, hexagonal, and triangular lattices once the dominant intermediates can be distributed across more GPUs.

All results in this section are estimated from fixed contraction paths produced by a state-of-the-art path finder.
In other words, the challenge is no longer only to find a good contraction path in the abstract, but to parallelize that path across multiple GPUs with hardware-efficient mode layouts, distribution choices, and communication schedules.
The next sections address this missing execution layer and show how our techniques convert such fixed paths into high-performance multi-GPU execution plans.

\section{From Contraction Paths to Multi-GPU Execution}
\label{sec:system}

\begin{figure}[htb]
  \centering
  \includegraphics[width=0.75\linewidth]{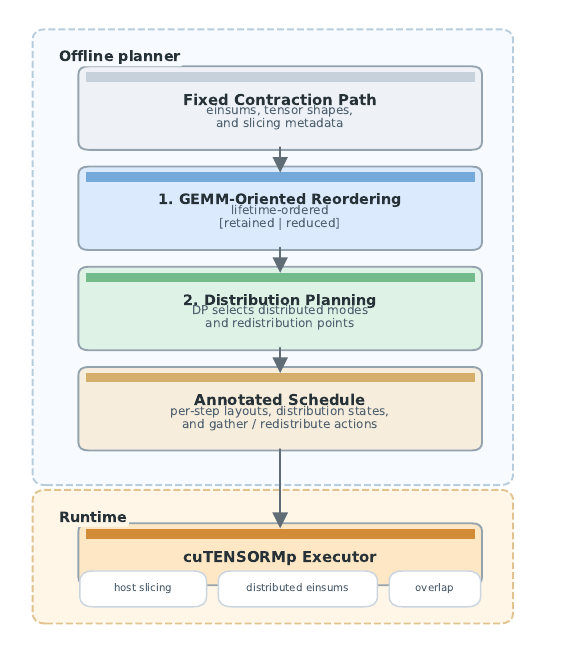}
  \caption{End-to-end workflow.
  The offline planner takes a fixed contraction path and applies two transformations: mode reordering produces GEMM-ready tensor layouts via lifetime ordering, and distribution planning selects distributed modes and redistribution points via DP search.
  The resulting annotated schedule is replayed at runtime by the cuTENSORMp executor with compute--communication overlap.}
  \label{fig:pipeline}
\end{figure}

We assume a fixed binary contraction path produced by an existing optimizer and focus on converting that path into an efficient multi-GPU execution.
Our framework performs two offline transformations---\emph{GEMM-Oriented Mode Reordering} and \emph{Communication-Aware Mode Distribution Planning}---and then replays the resulting annotated schedule with a cuTENSORMp-based executor.
Fig.~\ref{fig:pipeline} illustrates the end-to-end workflow.

The input is a contraction-path descriptor produced by an upstream path finder, containing the initial tensors, the pairwise contraction sequence, and any slicing metadata.
The \emph{planner} operates entirely offline: mode reordering first rewrites mode orders so that every pairwise contraction maps to a GEMM-like memory layout, then the distribution planner decides which tensor modes to partition across GPUs and when to redistribute or gather data.
The \emph{executor} replays the annotated schedule at runtime: one MPI rank per GPU loads its input, applies host-side slicing, and delegates each contraction step---including any inter-device communication---to cuTENSORMp~\cite{cutensormp}.

\subsection{GEMM-Oriented Mode Reordering}
\label{sec:mode-reordering}

Given a fixed contraction path of $S$ pairwise einsum equations, mode reordering permutes the modes of each tensor so that every pairwise contraction admits a regular, matrix-multiplication-like layout---without changing the contraction tree or the mathematical result.
Existing libraries such as cuTENSOR~\cite{cutensor} and TBLIS~\cite{matthews2018high} handle arbitrary mode orders at runtime for individual contractions, inserting transpose or permutation kernels as needed.
Mode reordering takes the complementary approach: it eliminates transposes \emph{ahead of time} by choosing globally consistent mode orders across the entire contraction tree.

Each pairwise contraction $A \times B \to C$ partitions its modes into \emph{retained} modes (surviving in~$C$) and \emph{reduced} modes (summed over).
Grouping the retained modes as the GEMM $m$/$n$ dimensions and the reduced modes as $k$, the canonical operand layout is $[\text{retained} \| \text{reduced}]$, which maps directly to a matrix multiplication.
The reordering pass seeks to realize this canonical structure at every step simultaneously.

\subsubsection{Backward pass and emergent lifetime ordering}

The key observation is that each mode in a contraction path has a well-defined \emph{remaining lifetime}: the number of contraction steps until it is summed over.
Reduced modes at step~$a$ have lifetime~0 (they are contracted at~$a$); retained modes survive to some later step and therefore have lifetime~$\ge 1$; open modes that persist to the final output have the longest lifetime.
A single backward pass over the contraction tree---from the last step to the first---propagates the $[\text{retained} \| \text{reduced}]$ constraint from each consumer to its producers:

\begin{enumerate}
  \item The output mode order of each step has already been set when the backward pass processed its downstream consumer.
  (The root's output order is fixed by the problem specification, e.g., the open indices of the TN.)
  \item For each input operand of step~$a$, extract the retained modes \emph{in the order they appear in the output} (preserving the ordering already established) and append the reduced modes.
  This yields the operand layout $[\text{retained in output order} \| \text{reduced}]$.
  \item The operand is the output of some earlier producer step~$p$.
  Apply the same permutation to the output of~$p$ so that the data layout stays consistent.
\end{enumerate}
Each producer is modified at most once (the first consumer to request it wins).

An important structural property emerges: after the backward pass, every tensor's modes are sorted by remaining lifetime---longest-lived modes leftmost, shortest-lived rightmost.
This is not an explicit sorting step; it falls out naturally from the recursive propagation of the $[\text{retained} \| \text{reduced}]$ constraint.
At each step, reduced modes (lifetime~0) are placed rightmost, and within the retained group, modes destined for later contraction are pre-sorted for their downstream consumers.
The result is a globally consistent layout in which:
\begin{enumerate}
  \item Every operand is in $[\text{retained} \| \text{reduced}]$ form, mapping directly to a GEMM call.
  \item Modes that survive many future steps sit in leading (leftmost) positions, while modes about to be contracted sit in trailing positions.
  \item Each downstream consumer receives its inputs already in the layout it expects, eliminating runtime transposes.
\end{enumerate}

\begin{figure}[htb]
  \centering
  \includegraphics[width=\linewidth]{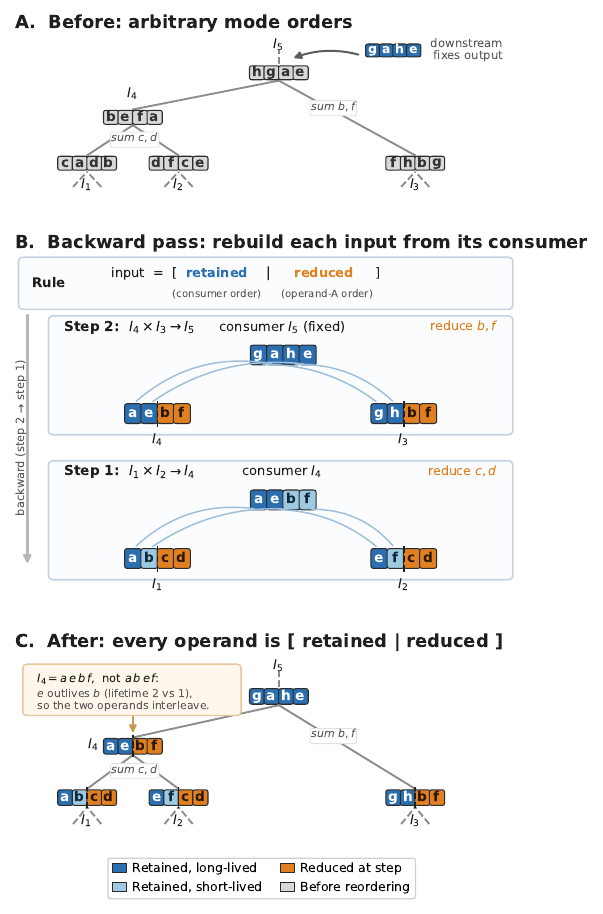}
  \caption{GEMM-oriented mode reordering on a two-step subtree.
  Dashed edges show connections to the rest of the contraction tree.
  Panel~A: arbitrary local orders plus the downstream request $I_5=\texttt{gahe}$.
  Panel~B: the backward pass visits step~2 and then step~1; at each visit, each input is rebuilt as ``shared modes in consumer order, then modes reduced here.''
  Panel~C: the resulting tree has $[\text{retained}\,|\,\text{reduced}]$ operand layouts (dashed split line), with modes shaded by remaining lifetime and reduction status.
  The output $I_4$ interleaves modes from $I_1$ and $I_2$ by lifetime, so step~2 can use $I_4$ as $[\text{retained}\,|\,\text{reduced}]$ without a transpose.}
  \label{fig:mode-reorder}
\end{figure}

Fig.~\ref{fig:mode-reorder} illustrates this on a two-step subtree of a larger contraction tree.
Dashed edges indicate that $I_1$--$I_3$ are outputs of earlier contractions and $I_5$ feeds into a later one; none are initial input tensors or the final result.
Panel~A shows arbitrary local mode orders (gray) together with the downstream-imposed order $I_5=\texttt{gahe}$, which seeds the backward pass.
Panel~B shows the actual rewrite rule applied at each visited contraction:
\[
\text{input} = [\text{shared in consumer order} \mid \text{reduced here}].
\]
Starting from the fixed output order of $I_5$, the pass first rewrites the step-2 inputs as $I_4=\texttt{ae|bf}$ and $I_3=\texttt{gh|bf}$, and then propagates the same consumer-imposed permutation backward to the producer outputs $I_1=\texttt{ab|cd}$ and $I_2=\texttt{ef|cd}$.
Panel~C shows the resulting operand layouts, where retained modes precede reduced modes, and the dashed split line marks the GEMM boundary for each contraction input.
Note the interleaved output of step~1: $I_4 = \texttt{aebf}$ rather than $\texttt{abef}$, because mode~$e$ (lifetime~2, from~$I_2$) outlives mode~$b$ (lifetime~1, from~$I_1$).
Thus an output order may interleave modes from the two operands; the invariant is that each downstream consumer receives its inputs in $[\text{retained}\,|\,\text{reduced}]$ form without any transpose.

The backward pass is deterministic: the contraction tree uniquely determines every mode's lifetime, so there is exactly one lifetime ordering and therefore one reordered scheme.
All transformations act exclusively on the ordering of modes within each tensor; no mode labels are created, deleted, or reassigned, and the mathematical result is unchanged.

\subsection{Communication-Aware Mode Distribution Planning}
\label{sec:distributed-modes}

Given the reordered contraction path and a target GPU count $P$, the distribution planner decides (i)~which contraction steps require multi-GPU execution, (ii)~which tensor modes to partition across devices at each such step, and (iii)~when to redistribute or gather data between steps.

\begin{figure}[htb]
  \centering
  \includegraphics[width=\linewidth]{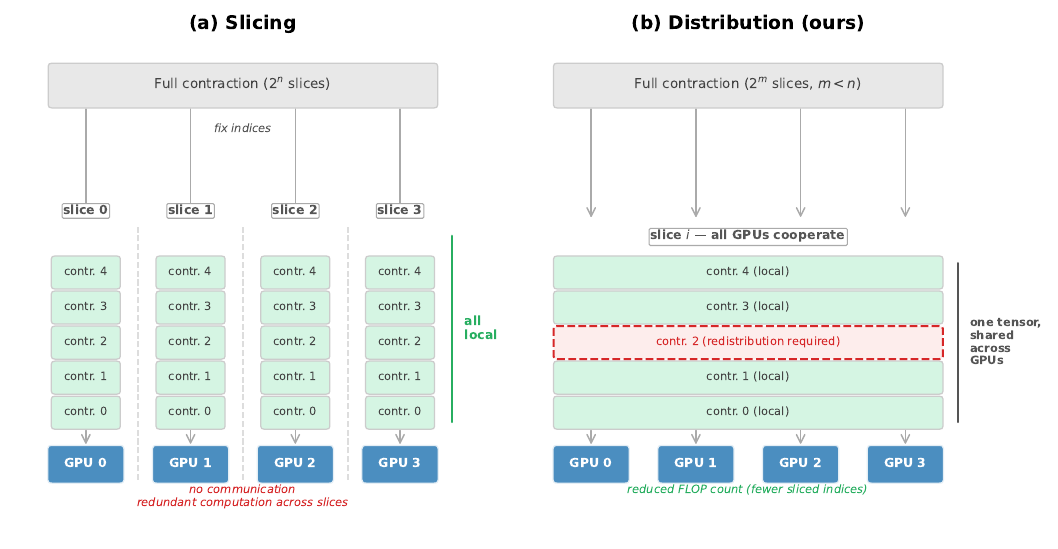}
  \caption{Slicing vs.\ distribution.
  (a)~Slicing fixes indices to create independent sub-tasks with no communication but redundant computation.
  (b)~Distribution partitions a single tensor across GPUs, reducing total FLOPs at the cost of communication that our planner minimizes.}
  \label{fig:slicing-vs-dist}
\end{figure}

This problem is distinct from slicing (Fig.~\ref{fig:slicing-vs-dist}).
Slicing eliminates a mode from the contraction entirely and creates independent sub-problems that share no data; slicing $n_{\mathrm{slice}}$ indices yields $2^{n_{\mathrm{slice}}}$ independent tasks that can be parallelized trivially but with redundant computation across tasks.
In contrast, distribution partitions one or more modes of an intermediate tensor across devices so that each device holds a shard of the same logical tensor, reducing the total FLOP count at the cost of inter-device communication whose volume must be controlled.

\subsubsection{Large Tensor Identification and Distribution}

We first identify the large intermediate tensors that drive the need for distribution.
A contraction step is considered \emph{large} if any of its operands or its output exceeds a configurable memory threshold~$s$ (8~GiB in our experiments):
\[
  \textproc{Large}(a) \;\equiv\;
  \max\bigl\{\lvert A^{(a)}\rvert,\;\lvert B^{(a)}\rvert,\;\lvert C^{(a)}\rvert\bigr\}
  \;\ge\; s.
\]
For each large tensor, we collect its \emph{use-chain}: the sequence of subsequent contractions in which the tensor or its descendants participate.
Any operand of a large contraction step may be distributed; cuTENSORMp supports distributing both operands and the output simultaneously.

When a large tensor is first distributed across $P$ devices, we select the minimum prefix of its leading modes whose product of dimensions is at least $P$:
\begin{equation}
  D = \arg\min_{|D'|} \left\{ D' \subseteq \text{prefix}(\alpha) \;\middle|\; \prod_{x \in D'} d_x \ge P \right\}.
  \label{eq:prefix}
\end{equation}
The modes in~$D$ are called the \emph{distributed modes}, and each device receives a contiguous shard of the tensor along these outermost dimensions.

This prefix-based distribution interacts directly with the GEMM-oriented mode reordering of Section~\ref{sec:mode-reordering}.
Because all tensors are stored in row-major order, the leading (leftmost) modes correspond to the outermost memory dimensions.
The backward pass sorts modes by remaining lifetime---longest-lived leftmost---so the leading prefix modes selected for distribution are precisely the modes that survive the most contraction steps.
This has two important consequences.
First, because the distributed modes are outermost in row-major layout, each device's shard is a single contiguous block in memory (no stride gaps), which maximizes the granularity of inter-device transfers and avoids the overhead of many small, non-contiguous messages.
Second, the distributed modes are stable across consecutive contractions: because they are the longest-lived, they tend not to be contracted away in the near term, reducing the frequency of forced redistributions.

\subsubsection{Distributed Mode Tracking}
\label{sec:tracking}

Once a tensor becomes distributed, its distributed modes must be tracked through subsequent contractions despite mode relabeling.
Each mode label in the contraction path is tagged with the equation at which it will be contracted (or marked as an open index if it survives to the final output).
At each step involving a distributed tensor, the planner identifies which of the current operand's modes correspond to the previously distributed modes by matching these tags, and propagates the distributed status accordingly.

Each step along a use-chain falls into one of four states:
\begin{itemize}
  \item \textbf{Activate}: the tensor is first distributed; prefix modes are selected per Eq.~\eqref{eq:prefix}.
  \item \textbf{Keep}: the output inherits the distributed modes from its operands; no communication is needed.
  \item \textbf{Redistribute}: fresh distributed modes are selected for the output, triggering an all-to-all data shuffle.
  A redistribution is \emph{structurally forced} when a currently distributed mode is about to be contracted at the next step; it may also be \emph{elective}, chosen by the planner to maintain large contiguous blocks per device.
  \item \textbf{Gather}: the tensor is small enough to fit on a single device; distributed modes are cleared and the data is replicated.
\end{itemize}
\subsubsection{Redistribution Point Selection via Dynamic Programming}
\label{sec:dp-search}

The key planning question is \emph{where} along each use-chain to place elective redistributions.
For a use-chain of $L$ equations, each presents a binary choice: \emph{keep} or \emph{redistribute}.
The raw search space is $2^L$, but we collapse it with dynamic programming by observing that the future cost depends only on which modes are currently distributed, not on the full history of past decisions.

\paragraph{Cost model}
Each transition from equation $a$ to $a{+}1$ incurs a cost composed of a local-compute term and, if redistributing, a communication term:
\begin{equation}
  t^{(a)} = t_{\mathrm{gemm}}^{(a)} + \mathbb{1}[\text{redist}]\;\cdot\;t_{\mathrm{comm}}^{(a)}.
\end{equation}
The GEMM time is estimated from per-device tensor sizes after distributing the selected modes:
\begin{equation}
  t_{\mathrm{gemm}} = \max\!\left(\frac{D_{\mathrm{rw}}}{B_{\mathrm{dev}}},\;\frac{\text{FLOPs}}{F_{\mathrm{dev}}}\right),
\end{equation}
where $D_{\mathrm{rw}}$ is the total bytes read and written per device, $B_{\mathrm{dev}}$ is the device memory bandwidth, and $F_{\mathrm{dev}}$ is the peak FLOP rate.

The redistribution time captures both bulk transfer and per-block overhead:
\begin{equation}
  t_{\mathrm{comm}} = \underbrace{\frac{|C^{(a)}|\,(P-1)}{P \cdot B_{\mathrm{net}}}}_{\text{bandwidth term}}
  \;+\;
  \underbrace{n_{\mathrm{blk}} \cdot \max\!\left(\lambda,\;\frac{s_{\mathrm{blk}}}{B_{\mathrm{net}}}\right)}_{\text{block-granularity term}},
  \label{eq:comm-cost}
\end{equation}
where $P$ is the GPU count, $B_{\mathrm{net}}$ the interconnect bandwidth, $\lambda$ the per-message latency, $n_{\mathrm{blk}}$ the number of contiguous blocks per device, and $s_{\mathrm{blk}}$ the base block size.
The block-granularity term naturally penalizes redistributions that produce many small messages (latency-bound) while treating large-block transfers as essentially free beyond their bandwidth cost.

A natural consequence of this cost model is that the DP prefers to redistribute at points where the tensor is small: redistributing a 16\,GB tensor costs ${\sim}16\times$ less than a 256\,GB tensor at the same step.
The DP thus automatically discovers that redistribution should concentrate at size valleys along the use-chain.

\paragraph{DP formulation}
Let $\mathcal{D}_a$ denote the set of distributed modes at the output of equation~$a$.
The DP state at equation~$a$ is simply $\mathcal{D}_a$.
At each equation, the planner propagates $\mathcal{D}_{a-1}$ to the current operand modes via mode-label tracking and evaluates two transitions: \emph{keep} (inherit the propagated modes, cost $t_{\mathrm{keep}}^{(a)}$) or \emph{redistribute} (select fresh leading prefix modes, cost $t_{\mathrm{redist}}^{(a)}$).
Because mode reordering has already sorted modes by remaining lifetime (Section~\ref{sec:mode-reordering}), the leading prefix modes are the longest-lived, making them the most stable choice for distribution and minimizing the frequency of forced redistributions.
When a distributed mode is nonetheless about to be contracted, redistribution is forced and only the redistribute transition is considered.
The DP selects the minimum-cost path through the use-chain:
\[
  \mathrm{dp}[a][\mathcal{D}_a] = \min_{\mathcal{D}_{a-1}} \bigl(\mathrm{dp}[a{-}1][\mathcal{D}_{a-1}] + t^{(a)}\bigr),
\]
with a gather cost added at the final equation.
Backtracing yields the optimal redistribution set~$R^\star$.

In practice, the state space is small.
When all distributed mode extents equal~2, the number of distributed modes is $\lceil\log_2 P\rceil$, and at each equation the DP tracks only the distinct sets of $\lceil\log_2 P\rceil$ modes that can occupy the leading prefix positions.
The total number of state evaluations across a typical use-chain is a few hundred, making the DP essentially instantaneous.

\paragraph{Why deferring redistribution can hurt}
A subtle failure mode motivates the DP over simpler heuristics.
If redistribution is deferred too long, the distributed modes may drift into unfavorable stride positions as subsequent contractions reshape the tensor.
When a forced redistribution finally becomes unavoidable---because a distributed mode is about to be contracted---it may occur at a point where the tensor is large (hundreds of gigabytes) and the stride pattern produces millions of tiny memory blocks, making the transfer latency-bound and extremely expensive.
The DP avoids this by redistributing proactively at points where tensors are small and block granularity is favorable, rather than waiting until a forced redistribution occurs at a costly peak.

\begin{figure}[htb]
  \centering
  \includegraphics[width=\linewidth]{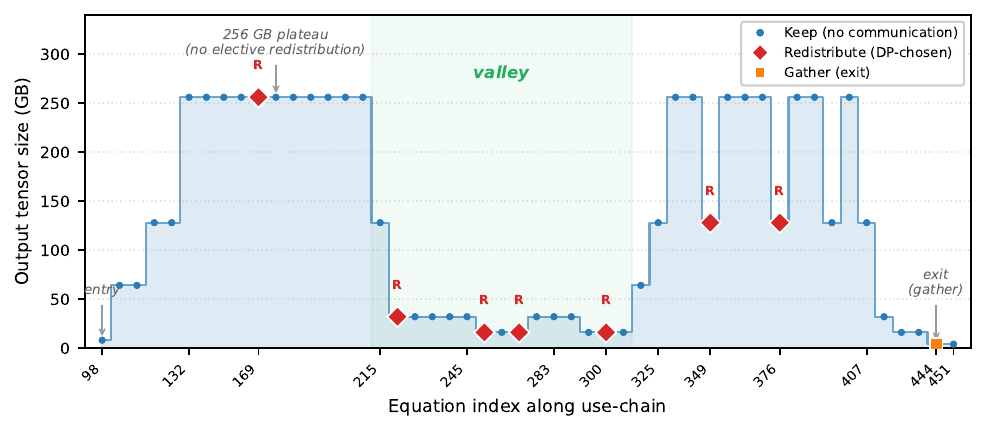}
  \caption{DP redistribution-point selection for the Zuchongzhi n60m24 benchmark on 8 H100 GPUs.
  The shaded area shows the distributed tensor's output size along the use-chain.
  Blue dots are equations where distributed modes are kept (no communication); red diamonds mark DP-chosen redistributions.
  The DP avoids redistributing on the 256\,GB plateau, instead concentrating transfers in the valley and at local minima of the second peak.}
  \label{fig:dp-redist}
\end{figure}

Fig.~\ref{fig:dp-redist} illustrates the DP result on the Zuchongzhi n60m24 benchmark with 8~GPUs.
The distributed tensor's output size varies from 4\,GB to 256\,GB across a 50-equation use-chain.
During the initial 256\,GB plateau (equations 132--205), no elective redistribution occurs---only one forced redistribution at equation~169, where the current distributed modes are about to be contracted.
The DP then places proactive redistributions in the valley (equations 223--300, where tensors are 16--32\,GB) and at local minima within the second peak (equations 349 and 376, both at 128\,GB).
The total redistributed volume is 596\,GB---only 4.6\% of the overall data movement---demonstrating that the DP concentrates communication at the cheapest points along the use-chain.

\section{Experiments}
\label{sec:experiments}

All experiments use DGX H100 nodes (Section~\ref{sec:hardware}).
All tensors use complex64 (single-precision complex) storage with FP32 arithmetic for contraction kernels.
For each GPU count, the path finder is run with a fixed time budget (approximately one hour) to produce a contraction path optimized for that configuration's memory capacity.
Because path optimization is NP-hard, the quality of the best path found within this budget varies across GPU counts.
Once a path is selected, it is held fixed and our framework applies mode reordering (Section~\ref{sec:mode-reordering}) and communication-aware distribution planning (Section~\ref{sec:distributed-modes}) on top of it.
Thus, the scaling results below measure the end-to-end, capacity-aware workflow: increasing the GPU count changes both the executable distributed schedule and the amount of slicing required by the selected path.
They should not be interpreted as fixed-path strong scaling; the non-monotonic trends below expose the remaining path-quality variation.

\begin{figure*}[t]
  \centering
  \includegraphics[width=\textwidth]{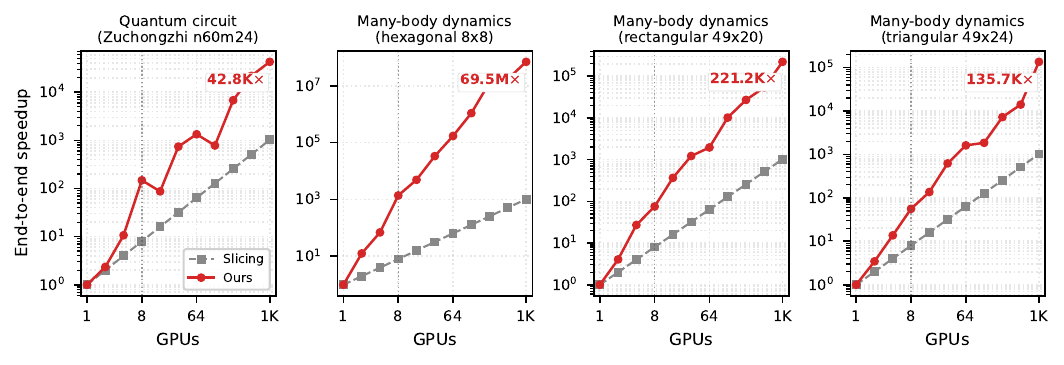}
  \caption{Projected full-contraction speedup vs.\ embarrassingly parallel slicing (dashed) on 1--1024 H100 GPUs, computed from measured per-slice runtime and sliced-bond count.
  The dotted vertical line marks one DGX H100 node (8 GPUs); larger configurations use InfiniBand across nodes.}
  \label{fig:scaling-4panel}
\end{figure*}

\subsection{Metrics}

For a configuration using $P$ GPUs, let $b_P$ denote the number of sliced bonds in the selected path and let $t_P$ denote the measured wall-clock runtime for one sliced subproblem executed by the distributed contraction engine.
When the full contraction contains $2^{b_P}$ independent slices, we report the projected full-contraction time
\begin{equation}
  T_P^{\mathrm{proj}} = t_P\,2^{b_P}.
\end{equation}
The projected full-contraction speedup relative to the one-GPU configuration is
\begin{equation}
  S_P = \frac{T_1^{\mathrm{proj}}}{T_P^{\mathrm{proj}}}
      = \frac{t_1\,2^{b_1}}{t_P\,2^{b_P}}.
  \label{eq:projected-speedup}
\end{equation}
We do not execute every slice when $2^{b_P}$ is very large; instead, we measure per-slice execution time and combine it with the path's sliced-bond count using the standard slicing cost model above.
This projection does not include external slice scheduling overhead or the final accumulation across slices.

The baseline is ideal embarrassingly parallel slicing, in which the contraction is decomposed into independent sub-tasks distributed evenly across GPUs with no inter-device communication.
In this regime, the speedup equals the GPU count: $P$ GPUs yield a $P\times$ speedup.
The \emph{extra speedup} is
\begin{equation}
  E_P = \frac{S_P}{P},
\end{equation}
quantifying the benefit of distributed contraction beyond ideal slicing.
The \emph{complexity reduction} is
\begin{equation}
  R_P = \frac{C_{t,1}}{C_{t,P}},
\end{equation}
where $C_{t,P}$ is the full-contraction FLOP count, including all slices, for the path selected at $P$ GPUs.

\subsection{Scaling Across Applications}

We evaluate four workloads: Zuchongzhi n60m24 quantum circuit simulation and three many-body dynamics benchmarks on hexagonal, rectangular, and triangular lattices.
The 1--8 GPU points run within a single DGX H100 NVLink domain; larger GPU counts span multiple DGX H100 nodes connected by InfiniBand.
Table~\ref{tab:single-node} reports the 8-GPU single-node point, and Fig.~\ref{fig:scaling-4panel} shows the full sweep from 1 to 1024 GPUs.

\begin{table}[htb]
  \centering
  \caption{Single-node results on 8 H100 GPUs (NVLink).
  Speedup is projected to the full contraction using Eq.~\eqref{eq:projected-speedup}.
  Extra speedup is projected speedup divided by the $8\times$ embarrassingly parallel baseline.
  Complexity reduction is the compute-only FLOP reduction from distributing intermediates (communication-free, hence an optimistic estimate rather than an achievable bound).}
  \label{tab:single-node}
  \small
  \setlength{\tabcolsep}{3pt}
  \begin{tabular}{lrrrr}
    \toprule
    Workload & \makecell{Full\\speedup} & \makecell{Extra\\speedup} & \makecell{Complexity\\reduction} & \makecell{TFLOP/s\\per GPU} \\
    \midrule
    Circuit (n60m24)         & $148\times$     & $18.5\times$  & $18.5\times$  & 28.1 \\
    Hexagonal 8$\times$8     & $1{,}383\times$ & $172.9\times$ & $197.8\times$ & 31.6 \\
    Rectangular 49$\times$20 & $75\times$      & $9.4\times$   & $9.3\times$   & 32.9 \\
    Triangular 49$\times$24  & $56\times$      & $7.0\times$   & $7.4\times$   & 31.8 \\
    \bottomrule
  \end{tabular}
\end{table}

At the 8-GPU NVLink point, all four workloads exhibit significant super-linear speedup beyond the $8\times$ slicing baseline, with extra speedups ranging from $7.0\times$ (triangular) to $172.9\times$ (hexagonal).
Because NVLink's high bandwidth (900\,GB/s per GPU) keeps the communication term small relative to compute, distribution converts essentially all of the available compute reduction into wall-clock gains: the achieved extra speedup reaches 87--101\% of the compute-only FLOP reduction, and all workloads sustain 28--33\,TFLOP/s per GPU.
To our knowledge, these are the first distributed multi-GPU TN contraction benchmarks for real-time quantum many-body dynamics on two-dimensional lattices.

\subsection{Multi-Node Scaling to 1024 GPUs}

We next scale all four workloads to 1024 GPUs across multiple DGX H100 nodes connected by InfiniBand, where inter-node bandwidth is substantially lower than intra-node NVLink.
Table~\ref{tab:multi-node-apps} summarizes the 1024-GPU results.

\begin{table}[htb]
  \centering
  \caption{Multi-node results on 1024 H100 GPUs over InfiniBand.
  Runtime is measured per sliced subproblem; projected full speedup combines slicing parallelism and distributed contraction.
  Extra speedup isolates the gain beyond the $1024\times$ embarrassingly parallel baseline and is our primary metric; complexity reduction is the compute-only FLOP reduction (communication-free), not an achievable bound.}
  \label{tab:multi-node-apps}
  \small
  \setlength{\tabcolsep}{2.5pt}
  \resizebox{\linewidth}{!}{%
  \begin{tabular}{lrrrrr}
    \toprule
    Workload & \makecell{Per-slice\\runtime (s)} & \makecell{Sliced\\bonds} & \makecell{Projected full\\speedup} & \makecell{Extra\\speedup} & \makecell{Complexity\\reduction} \\
    \midrule
    Circuit (n60m24)         & 20.19  & 20 & $42{,}759\times$     & $41.8\times$      & $418\times$ \\
    Hexagonal 8$\times$8     & 113.27 & 6  & $69.5$M$\times$      & $67{,}869\times$  & $1.49$M$\times$ \\
    Rectangular 49$\times$20 & 34.70  & 14 & $221{,}212\times$    & $216.0\times$     & $3{,}154\times$ \\
    Triangular 49$\times$24  & 12.19  & 14 & $135{,}728\times$    & $132.6\times$     & $986\times$ \\
    \bottomrule
  \end{tabular}
  }
\end{table}

At 1024 GPUs, all four workloads remain well above the linear slicing baseline.
We take the \emph{extra speedup}---the gain beyond the $1024\times$ embarrassingly parallel baseline---as the primary, technique-attributable metric: it ranges from $41.8\times$ on the quantum-circuit benchmark to $67{,}869\times$ on the hexagonal many-body benchmark, the latter driven by a reduction in sliced bonds from 37 to 6.
The corresponding projected full-contraction speedups in Table~\ref{tab:multi-node-apps} are far larger because they also fold in the path's slicing parallelism; since that factor is governed by the path finder and the available memory rather than by our execution layer, we report it for completeness but treat the extra speedup as the headline result.

\subsubsection{Non-monotonic scaling}
Neither the extra speedup nor the complexity reduction increases monotonically with GPU count.
For example, on Zuchongzhi n60m24, the extra speedup reaches $18.49\times$ at 8 GPUs, drops to $5.44\times$ at 16 GPUs, and later recovers to $41.76\times$ at 1024 GPUs; similarly, the complexity reduction peaks at $522.54\times$ at 512 GPUs but falls to $418.03\times$ at 1024 GPUs.
Two factors contribute to this non-monotonicity.
First, contraction-path quality varies across GPU counts: because path optimization is NP-hard, we allocate the same fixed time budget (approximately one hour) to the path finder for each configuration, and the best path found within that budget can differ substantially---this directly affects the complexity reduction, which depends solely on the path.
Second, 8 GPUs reside within a single NVLink domain on DGX H100, where redistribution is fast and cheap; at 16 GPUs, the execution spans two nodes connected by InfiniBand, introducing higher-latency cross-node communication that further reduces the achieved extra speedup.

Notably, even when scaling beyond the NVLink domain onto InfiniBand interconnect, the framework continues to deliver substantial extra speedups.
At 1024 GPUs, the extra speedup ranges from $41.8\times$ on the quantum-circuit workload to $67{,}869\times$ on the hexagonal many-body workload, demonstrating that communication-aware distributed contraction remains highly effective even over lower-bandwidth inter-node networks.

\subsubsection{Communication, not arithmetic, bounds the achieved speedup}
The \emph{complexity reduction} column in Table~\ref{tab:multi-node-apps} reports the compute-only FLOP reduction from distributing intermediate tensors.
It deliberately excludes communication and is therefore not an achievable performance bound: distribution is fundamentally a compute--communication trade, and on InfiniBand the communication term is substantial.
The meaningful comparison is thus between interconnect regimes, not against this communication-free estimate.
Within a single NVLink node, where inter-GPU bandwidth is high, the extra speedup tracks the compute-only reduction almost exactly (87--101\%).
Across nodes over InfiniBand, redistribution, cross-node collectives, and broadcast costs raise the communication term, yet the extra speedup remains large in absolute terms ($41.8$--$67{,}869\times$).
In short, the benefit of distribution is limited by communication rather than by arithmetic, and it grows with interconnect bandwidth.

\section{Conclusion}
\label{sec:conclusion}

We presented a multi-GPU framework for exact TN contraction that moves beyond embarrassingly parallel slicing.
The two core techniques---GEMM-oriented mode reordering and communication-aware mode distribution planning---are fully general: they depend only on the contraction tree, mode lifetimes, and tensor shapes, not on any application-specific property.
A compute-only analysis across six diverse workloads---quantum circuit simulation, quantum error correction, combinatorial optimization on King's subgraphs, and many-body dynamics on rectangular, hexagonal, and triangular lattices (Fig.~\ref{fig:complexity-reduction})---shows that the FLOP reduction from distributed memory is consistent across application families.
Per-slice execution measurements on four of these workloads sustain 28--33\,TFLOP/s per GPU (42--49\% of H100 FP32 peak).
Within a single NVLink node, distribution yields $7.0$--$173\times$ extra speedup beyond slicing, capturing 87--101\% of the compute-only reduction because NVLink's high intra-node bandwidth leaves communication well below the compute time.

Across 1024 H100 GPUs over InfiniBand---where inter-node bandwidth is an order of magnitude lower than NVLink---the extra speedup beyond embarrassingly parallel slicing remains large: $41.8\times$ on Zuchongzhi n60m24, $67{,}869\times$ on hexagonal many-body dynamics, $216\times$ on rectangular many-body dynamics, and $132.6\times$ on triangular many-body dynamics.

These results establish a clear principle: distribution trades arithmetic for communication, so its benefit is limited by interconnect bandwidth rather than by the available FLOP reduction.
On NVLink, where bandwidth is high, this communication cost is small and the extra speedup tracks the compute-only reduction; on InfiniBand it becomes the dominant cost---yet in both regimes the extra speedup stays far above linear slicing.
This makes the framework particularly well-suited to emerging platforms such as the NVIDIA GB200 NVL72, which extends NVLink-class bandwidth to 72 GPUs (13.8\,TB aggregate memory, 1.8\,TB/s per GPU), and future multi-node NVLink (MNNVL) systems that further widen the high-bandwidth domain.
As interconnect technology continues to advance, a larger share of the compute reduction will translate into wall-clock gains, making communication-aware distributed contraction an increasingly powerful alternative to slicing for frontier TNs.

\section*{Disclosure}
AI writing assistants (Claude Code and Codex) were used for drafting and refining text, generating figure-plotting scripts, and restructuring sections.
All technical content, experimental data, algorithm design, and scientific claims originate from the authors.
All AI-generated output was reviewed and validated by the authors.

\bibliographystyle{IEEEtran}
\bibliography{ref}

@article{markov2008simulating,
  title     = {Simulating quantum computation by contracting tensor networks},
  author    = {Markov, Igor L and Shi, Yaoyun},
  journal   = {SIAM Journal on Computing},
  volume    = {38},
  number    = {3},
  pages     = {963--981},
  year      = {2008},
  publisher = {SIAM}
}

@article{arute2019quantum,
  title     = {Quantum supremacy using a programmable superconducting processor},
  author    = {Arute, Frank and Arya, Kunal and Babbush, Ryan and Bacon, Dave and Bardin, Joseph C and Barends, Rami and Biswas, Rupak and Boixo, Sergio and Brandao, Fernando GSL and Buell, David A and others},
  journal   = {Nature},
  volume    = {574},
  number    = {7779},
  pages     = {505--510},
  year      = {2019},
  publisher = {Nature Publishing Group},
  url       = {https://doi.org/10.1038/s41586-019-1666-5}
}

@article{wu2021strong,
  title     = {Strong quantum computational advantage using a superconducting quantum processor},
  author    = {Wu, Yulin and Bao, Wan-Su and Cao, Sirui and Chen, Fusheng and Chen, Yu and Chen, Xiawei and Chung, Tung-Hsun and Deng, Hui and Du, Yibo and Fan, Daojin and others},
  journal   = {Physical Review Letters},
  volume    = {127},
  number    = {18},
  pages     = {180501},
  year      = {2021},
  publisher = {APS},
  url       = {https://doi.org/10.1103/PhysRevLett.127.180501}
}

@article{gray2021hyper,
  title     = {Hyper-optimized tensor network contraction},
  author    = {Gray, Johnnie and Kourtis, Stefanos},
  journal   = {Quantum},
  volume    = {5},
  pages     = {410},
  year      = {2021},
  publisher = {Verein zur F{\"o}rderung des Open Access Publizierens in den Quantenwissenschaften},
  url       = {https://doi.org/10.22331/q-2021-03-15-410}
}

@article{schollwock2011dmrg,
  title     = {The density-matrix renormalization group in the age of matrix product states},
  author    = {Schollw{\"o}ck, Ulrich},
  journal   = {Annals of Physics},
  volume    = {326},
  number    = {1},
  pages     = {96--192},
  year      = {2011},
  publisher = {Elsevier},
  url       = {https://doi.org/10.1016/j.aop.2010.09.012}
}

@article{orus2014practical,
  title     = {A practical introduction to tensor networks: Matrix product states and projected entangled pair states},
  author    = {Or{\'u}s, Rom{\'a}n},
  journal   = {Annals of Physics},
  volume    = {349},
  pages     = {117--158},
  year      = {2014},
  publisher = {Elsevier},
  url       = {https://doi.org/10.1016/j.aop.2014.06.013}
}

@inproceedings{novikov2015tensorizing,
  title     = {Tensorizing Neural Networks},
  author    = {Novikov, Alexander and Podoprikhin, Dmitry and Osokin, Anton and Vetrov, Dmitry P.},
  booktitle = {Advances in Neural Information Processing Systems 28},
  pages     = {442--450},
  year      = {2015}
}

@inproceedings{stoudenmire2016supervised,
  title     = {Supervised Learning with Tensor Networks},
  author    = {Stoudenmire, E. Miles and Schwab, David J.},
  booktitle = {Advances in Neural Information Processing Systems 29},
  pages     = {4799--4807},
  year      = {2016}
}

@article{bravyi2014efficient,
  title     = {Efficient algorithms for maximum likelihood decoding of quantum error-correcting codes},
  author    = {Bravyi, Sergey and Suchara, Martin and Vargo, Alexander},
  journal   = {Physical Review A},
  volume    = {90},
  number    = {3},
  pages     = {032326},
  year      = {2014},
  publisher = {APS},
  url       = {https://doi.org/10.1103/PhysRevA.90.032326}
}

@article{ferris2014tensor,
  title     = {Tensor networks and quantum error correction},
  author    = {Ferris, Andrew J and Poulin, David},
  journal   = {Physical review letters},
  volume    = {113},
  number    = {3},
  pages     = {030501},
  year      = {2014},
  publisher = {APS},
  url       = {https://doi.org/10.1103/PhysRevLett.113.030501}
}

@article{fowler2012surface,
  title     = {Surface codes: Towards practical large-scale quantum computation},
  author    = {Fowler, Austin G and Mariantoni, Matteo and Martinis, John M and Cleland, Andrew N},
  journal   = {Physical Review A},
  volume    = {86},
  number    = {3},
  pages     = {032324},
  year      = {2012},
  publisher = {APS},
  url       = {https://doi.org/10.1103/PhysRevA.86.032324}
}

@article{google2023suppressing,
  title     = {Suppressing quantum errors by scaling a surface code logical qubit},
  author    = {{Google Quantum AI}},
  journal   = {Nature},
  volume    = {614},
  number    = {7949},
  pages     = {676--681},
  year      = {2023},
  publisher = {Nature Publishing Group},
  url       = {https://doi.org/10.1038/s41586-022-05434-1}
}

@article{liu2021tropical,
  title     = {Tropical Tensor Network for Ground States of Spin Glasses},
  author    = {Liu, Jin-Guo and Wang, Lei and Zhang, Pan},
  journal   = {Physical Review Letters},
  volume    = {126},
  number    = {9},
  pages     = {090506},
  year      = {2021},
  publisher = {APS},
  url       = {https://doi.org/10.1103/PhysRevLett.126.090506}
}

@article{suzuki1976generalized,
  title     = {Generalized Trotter's formula and systematic approximants of exponential operators and inner derivations with applications to many-body problems},
  author    = {Suzuki, Masuo},
  journal   = {Communications in Mathematical Physics},
  volume    = {51},
  number    = {2},
  pages     = {183--190},
  year      = {1976},
  publisher = {Springer},
  url       = {https://doi.org/10.1007/BF01609348}
}

@article{vidal2004efficient,
  title     = {Efficient simulation of one-dimensional quantum many-body systems},
  author    = {Vidal, Guifr{\'e}},
  journal   = {Physical Review Letters},
  volume    = {93},
  number    = {4},
  pages     = {040502},
  year      = {2004},
  publisher = {APS},
  url       = {https://doi.org/10.1103/PhysRevLett.93.040502}
}

@article{pan2020contracting,
  title     = {Contracting Arbitrary Tensor Networks: General Approximate Algorithm and Applications in Graphical Models and Quantum Circuit Simulations},
  author    = {Pan, Feng and Zhou, Pengfei and Li, Sujie and Zhang, Pan},
  journal   = {Physical Review Letters},
  volume    = {125},
  number    = {6},
  pages     = {060503},
  year      = {2020},
  publisher = {APS},
  url       = {https://doi.org/10.1103/PhysRevLett.125.060503}
}

@article{pan2022bigbatch,
  title     = {Simulation of Quantum Circuits Using the Big-Batch Tensor Network Method},
  author    = {Pan, Feng and Zhang, Pan},
  journal   = {Physical Review Letters},
  volume    = {128},
  number    = {3},
  pages     = {030501},
  year      = {2022},
  publisher = {APS},
  url       = {https://doi.org/10.1103/PhysRevLett.128.030501}
}

@article{pan2022sampling,
  title     = {Solving the Sampling Problem of the {Sycamore} Quantum Circuits},
  author    = {Pan, Feng and Chen, Keyang and Zhang, Pan},
  journal   = {Physical Review Letters},
  volume    = {129},
  number    = {9},
  pages     = {090502},
  year      = {2022},
  publisher = {APS},
  url       = {https://doi.org/10.1103/PhysRevLett.129.090502}
}

@article{pan2024gpus,
  title     = {Efficient Quantum Circuit Simulation by Tensor Network Methods on Modern {GPU}s},
  author    = {Pan, Feng and Gu, Hanfeng and Kuang, Lvlin and Liu, Bing and Zhang, Pan},
  journal   = {ACM Transactions on Quantum Computing},
  volume    = {5},
  number    = {4},
  year      = {2024},
  publisher = {ACM},
  url       = {https://doi.org/10.1145/3696465}
}

@article{huang2021parallelization,
  title     = {Efficient parallelization of tensor network contraction for simulating quantum computation},
  author    = {Huang, Cupjin and Zhang, Fang and Newman, Michael and Ni, Xiaotong and Ding, Dawei and Cai, Junjie and Gao, Xun and Wang, Tenghui and Wu, Feng and Zhang, Gengyan and Ku, Hsiang-Sheng and Tian, Zhengxiong and Wu, Junyin and Xu, Haihong and Yu, Huanjun and Yuan, Bo and Szegedy, Mario and Shi, Yaoyun and Zhao, Hui-Hai and Deng, Chunqing and Chen, Jianxin},
  journal   = {Nature Computational Science},
  volume    = {1},
  pages     = {578--587},
  year      = {2021},
  publisher = {Nature Publishing Group},
  url       = {https://doi.org/10.1038/s43588-021-00119-7}
}

@inproceedings{liu2021closing,
  title     = {Closing the ``Quantum Supremacy'' Gap: Achieving Real-Time Simulation of a Random Quantum Circuit Using a New {Sunway} Supercomputer},
  author    = {Liu, Yong and Liu, Xin and Li, Fang and Yang, Yuling and Song, Jiawei and Zhao, Pengpeng and Wang, Zhen and Peng, Dajia and Fu, Haohuan and Chen, Dexun and Wu, Wenzhao and Huang, Heliang and Guo, Chu},
  booktitle = {Proceedings of the International Conference for High Performance Computing, Networking, Storage and Analysis},
  pages     = {1--12},
  year      = {2021},
  publisher = {ACM},
  url       = {https://doi.org/10.1145/3458817.3487399}
}

@inproceedings{fu2024surpassing,
  title     = {Surpassing {Sycamore}: Achieving Energetic Superiority Through System-Level Circuit Simulation},
  author    = {Fu, Rong and Su, Zhongling and Zhong, Han-Sen and Zhao, Xian-He and Zhang, Jianyang and Pan, Feng and others},
  booktitle = {Proceedings of the International Conference for High Performance Computing, Networking, Storage and Analysis},
  year      = {2024},
  publisher = {IEEE},
  url       = {https://doi.org/10.1109/SC41406.2024.00085}
}

@article{zhao2025leapfrogging,
  title     = {Leapfrogging {Sycamore}: Harnessing 1432 {GPU}s for 7x Faster Quantum Random Circuit Sampling},
  author    = {Zhao, Xian-He and Zhong, Han-Sen and Pan, Feng and others},
  journal   = {National Science Review},
  volume    = {12},
  number    = {3},
  pages     = {nwae317},
  year      = {2025},
  publisher = {Oxford University Press},
  url       = {https://doi.org/10.1093/nsr/nwae317}
}

@article{zlokapa2023boundaries,
  title     = {The computational boundaries of quantum advantage},
  author    = {Zlokapa, Alexander and Fuchs, Falk and Schaeffer, Luke and Dalzell, Alexander M. and Lau, Edwin and Holland, Etienne T. and others},
  journal   = {npj Quantum Information},
  volume    = {9},
  pages     = {36},
  year      = {2023},
  publisher = {Nature Publishing Group},
  url       = {https://doi.org/10.1038/s41534-023-00744-7}
}

@article{piveteau2024beyond2d,
  title     = {Tensor-Network Decoding Beyond 2D},
  author    = {Piveteau, Christophe and Chubb, Christopher T. and Renes, Joseph M.},
  journal   = {PRX Quantum},
  volume    = {5},
  number    = {4},
  pages     = {040303},
  year      = {2024},
  publisher = {APS},
  url       = {https://doi.org/10.1103/PRXQuantum.5.040303}
}

@article{bausch2024learning,
  title     = {Learning high-accuracy error decoding for quantum processors},
  author    = {Bausch, Johannes and Kesselring, Markus S. and Elben, Andreas and Swaroop, Vedika and Yao, Boxi and Molle, Adrian and others},
  journal   = {Nature},
  volume    = {635},
  pages     = {834--840},
  year      = {2024},
  publisher = {Nature Publishing Group},
  url       = {https://doi.org/10.1038/s41586-024-08148-8}
}

@article{google2025belowthreshold,
  title     = {Quantum error correction below the surface code threshold},
  author    = {{Google Quantum AI} and others},
  journal   = {Nature},
  volume    = {638},
  pages     = {920--926},
  year      = {2025},
  publisher = {Nature Publishing Group},
  url       = {https://doi.org/10.1038/s41586-024-08449-y}
}

@article{liang2025kings,
  title         = {Independent Set Enumeration in King's Graphs by Tensor Network Contractions},
  author        = {Liang, Kai},
  journal       = {arXiv preprint arXiv:2505.12776},
  year          = {2025},
  archivePrefix = {arXiv},
  eprint        = {2505.12776},
  primaryClass  = {math.CO},
  url           = {https://arxiv.org/abs/2505.12776}
}

@article{fauseweh2024manybody,
  title     = {The state of quantum computing applications in challenging many-body quantum dynamics},
  author    = {Fauseweh, Benedikt},
  journal   = {Nature Communications},
  volume    = {15},
  pages     = {2123},
  year      = {2024},
  publisher = {Nature Publishing Group},
  url       = {https://doi.org/10.1038/s41467-024-46402-9}
}

@article{kim2023utility,
  title     = {Evidence for the utility of quantum computing before fault tolerance},
  author    = {Kim, Youngseok and Eddins, Andrew and Anand, Siddharth and Wei, Ken Xuan and van den Berg, Ewout and Rosenblatt, Sami and Nayfeh, Hasan and Wu, Yantao and Zaletel, Michael and Temme, Kristan and Kandala, Abhinav},
  journal   = {Nature},
  volume    = {618},
  pages     = {500--505},
  year      = {2023},
  publisher = {Nature Publishing Group},
  url       = {https://doi.org/10.1038/s41586-023-06096-3}
}

@article{shtanko2025integrability,
  title     = {Uncovering local integrability in quantum many-body dynamics},
  author    = {Shtanko, Oleksandr and Wang, Derek S. and Zhang, Haimeng and Harle, Nikhil and Seif, Alireza and Movassagh, Ramis and Minev, Zlatko},
  journal   = {Nature Communications},
  year      = {2025},
  publisher = {Nature Publishing Group},
  url       = {https://doi.org/10.1038/s41467-025-57623-x}
}

@article{king2023spinglass,
  title     = {Quantum critical dynamics in a 5,000-qubit programmable spin glass},
  author    = {King, Andrew D. and others},
  journal   = {Nature},
  volume    = {617},
  pages     = {61--66},
  year      = {2023},
  publisher = {Nature Publishing Group}
}

@article{king2025beyondclassical,
  title     = {Beyond-classical computation in quantum simulation},
  author    = {King, Andrew D. and Nocera, Alberto and Rams, Marek and Dziarmaga, Jacek and Wiersema, Roeland and others},
  journal   = {Science},
  pages     = {199--204},
  year      = {2025},
  publisher = {AAAS},
  url       = {https://doi.org/10.1126/science.ado6285}
}

@article{mildenberger2025confinement,
  title     = {Confinement in a {Z2} lattice gauge theory on a quantum computer},
  author    = {Mildenberger, Julius and Jiang, Zhang and Mruczkiewicz, Wojtek and Halimeh, Jad C. and Hauke, Philipp},
  journal   = {Nature Physics},
  year      = {2025},
  publisher = {Nature Publishing Group},
  url       = {https://doi.org/10.1038/s41567-024-02723-6}
}

@misc{cutensormp,
  title        = {{cuTENSORMp}: Multi-Process Tensor Contraction Library},
  author       = {{NVIDIA Corporation}},
  howpublished = {\url{https://docs.nvidia.com/cuda/cutensor/latest/user_guide_cutensorMp.html}},
  year         = {2024}
}

@misc{cutensor,
  title        = {{cuTENSOR}: A High-Performance {CUDA} Library for Tensor Primitives},
  author       = {{NVIDIA Corporation}},
  howpublished = {\url{https://developer.nvidia.com/cutensor}},
  year         = {2024}
}

@misc{nvidia_h100,
  title        = {{NVIDIA H100 Tensor Core GPU Architecture}},
  author       = {{NVIDIA Corporation}},
  howpublished = {\url{https://resources.nvidia.com/en-us-tensor-core}},
  year         = {2022}
}

@misc{nccl,
  title        = {{NCCL}: {NVIDIA} Collective Communications Library},
  author       = {{NVIDIA Corporation}},
  howpublished = {\url{https://developer.nvidia.com/nccl}},
  year         = {2024}
}

@article{brown2025multigpu,
  title   = {Multi-{GPU} Quantum Circuit Simulation and the Impact of Network Performance},
  author  = {Brown, W. Michael and Ramesh, Anurag and Lubinski, Thomas and Nguyen, Thien and Neira, David E. Bernal},
  journal = {arXiv preprint arXiv:2511.14664},
  year    = {2025},
  url     = {https://arxiv.org/abs/2511.14664}
}

@article{matthews2018high,
  title     = {High-Performance Tensor Contraction without Transposition},
  author    = {Matthews, Devin A.},
  journal   = {SIAM Journal on Scientific Computing},
  volume    = {40},
  number    = {1},
  pages     = {C1--C24},
  year      = {2018},
  url       = {https://doi.org/10.1137/16M108968X}
}

@inproceedings{xu2024atlas,
  title     = {Atlas: Hierarchical Partitioning for Quantum Circuit Simulation on {GPUs}},
  author    = {Xu, Mingkuan and Cao, Shiyi and Miao, Xupeng and Acar, Umut A. and Jia, Zhihao},
  booktitle = {Proceedings of the International Conference for High Performance Computing, Networking, Storage and Analysis (SC24)},
  year      = {2024},
  url       = {https://doi.org/10.1109/SC41406.2024.00072}
}

@article{cicero2025simulation,
  title     = {Simulation of Quantum Computers: Review and Acceleration Opportunities},
  author    = {Cicero, Alessio and Maleki, Mohammad Ali and Azhar, Muhammad Waqar and Kockum, Anton Frisk and Trancoso, Pedro},
  journal   = {ACM Transactions on Quantum Computing},
  volume    = {7},
  number    = {1},
  pages     = {3},
  year      = {2025},
  publisher = {ACM},
  url       = {https://doi.org/10.1145/3701725}
}

@article{chen2025swtnc,
  title   = {{SW-TNC}: Reaching the Most Complex Random Quantum Circuit via Tensor Network Contraction},
  author  = {Chen, Yaojian and Sun, Zhaoqi and Qiu, Chengyu and Li, Zegang and Liu, Yanfei and Gan, Lin and Duan, Xiaohui and Yang, Guangwen},
  journal = {arXiv preprint arXiv:2504.09186},
  year    = {2025},
  url     = {https://arxiv.org/abs/2504.09186}
}
\end{document}